\documentclass{aip-cp}

\usepackage[sort&compress,numbers]{natbib}
\usepackage{rotating}
\usepackage{graphicx}

\usepackage{enumerate}

\newcommand{\be}{\begin{equation}}
\newcommand{\ee}{\end{equation}}
\newcommand{\bea}{\begin{eqnarray}}
\newcommand{\eea}{\end{eqnarray}}
\newcommand{\eqref}[1]{(\ref{#1})}

\newcommand{\frw}{{\mbox{\tiny FRW}}}
\newcommand{\osc}{{\mbox{\tiny osc}}}

\newcommand{\gsim}{\lower.7ex\hbox{$\;\stackrel{\textstyle>}{\sim}\;$}}
\newcommand{\lsim}{\lower.7ex\hbox{$\;\stackrel{\textstyle<}{\sim}\;$}}

\newcommand{\beq}{\begin{equation}}
\newcommand{\eeq}{\end{equation}}

\begin{document}

% Title portion
\title{Thermal History of the Universe After Inflation}

\author[aff1]{Scott Watson\corref{cor1}}

\affil[aff1]{Department of Physics, Syracuse University, Syracuse, NY 13244, USA.}
\corresp[cor1]{Corresponding author: gswatson@syr.edu}

\maketitle

\begin{abstract}
When did the universe thermalize?  In this talk I review the status of this issue and its importance in
establishing the expected properties of dark matter, the growth of large-scale structure, and the viability of inflation models when confronted with CMB observations. I also present a novel approach to tackling the theoretical challenges surrounding inflationary (p)reheating, which seeks to extend past work on the Effective Field Theory of Inflation to the time of reheating.
\end{abstract}

% Head 1
\section{INTRODUCTION}
The success of Big Bang cosmology in predicting the correct abundances of the light elements and the existence of the Cosmic Microwave Background (CMB)  rests upon the assumption that the universe began in a hot, dense state and as the universe expanded it cooled.  This {\it thermal history} for the early universe can then be used to try to understand the primordial origin of dark matter (DM) particles,  baryons, and the matter / anti-matter asymmetry.  This picture is simple, elegant, and robust, but it can not be complete.  Indeed, the primordial density perturbations -- indirectly detected through the temperature anisotropies of the CMB and required for the growth of the large-scale structure of the universe -- can not be causally generated in a strictly thermal universe.  Instead, inflation provides the most compelling model for their origin, and during this epoch the universe was not in thermal equilibrium. Moreover, following inflation the universe must {\it reheat} providing the hot big bang. This process can occur far from equilibrium and result in an effectively matter dominated epoch that could last up to the time of Big Bang Nucleosynthesis (BBN).  Moreover, constructing self-consistent inflationary models typically requires invoking additional symmetries (such as Supersymmetry) and also requires the presence of additional scalar fields (moduli) with gravitational strength couplings. When these fields coherently oscillate about their low energy minimum they can come to dominate the post-inflationary expansion and provide another motivation for a matter dominated universe prior to BBN.  Moduli masses determined by symmetry breaking slightly above the TeV-Scale implies matter domination until shortly before BBN.  Thus, prolonged inflationary reheating and moduli in the early universe are both compelling theoretical motivations for taking seriously a matter dominated universe prior to BBN.  In this talk we will explore the observational implications for such a {\em non-thermal} history.

\section{POST-INFLATIONARY HISTORY AND OBSERVATIONAL IMPLICATIONS}
Whether prolonged inflationary reheating or moduli domination after inflation leads to a non-thermal history, this departure from a strictly radiation dominated universe leads to a number of interesting observational predictions. 
I will first consider the effects of this history on the production and expected microscopic properties of DM WIMPs, the growth of structure during this phase, and then implications for CMB observations and inflation. There are other phenomenological implications e.g. baryogenesis, axion DM, isocurvature modes in the CMB, and primordial black hole production, but given the time constraints I refer you to the recent review \cite{Kane:2015jia} for a more thorough discussion of both the theoretical motivation and the phenomenological consequences of a non-thermal history.  

\subsection{Dark Matter WIMPs}
The entropy production associated with the end of the non-thermal phase (whether from late-time reheating or moduli decay) dilutes any previous population of DM by $\Omega_{cdm} \rightarrow \Omega_{cdm} (T_r/T_f)^3$.  If WIMPs represent all of the DM then these particles must be produced again at the time of reheating, or their initial abundance must large enough that this suppression leads to the correct relic density.  As an example, for SUSY WIMPs non-thermal production of Winos or Higgsinos is an example of the former \cite{Moroi:1999zb}, whereas the Bino provides a viable candidate for the latter \cite{Erickcek:2015bda}.  What is interesting in both cases is that these particles can have self-annihilation rates that differ from strictly thermal relics and this leads to new predictions for the expected microscopic properties of DM. 
To see this consider the relic density in non-thermal DM resulting from scalar decays \cite{Acharya:2008bk,Acharya:2009zt}
\be  \label{nt1}
\Omega_{dm}^{NT}h^2 \simeq 0.10 \, \left( \frac{m_X}{100 \; \mbox{GeV}}\right) \left(\frac{10.75}{g_\ast} \right)^{1/4} \left( \frac{3 \times 10^{-24} \; \mbox{cm}^3/\mbox{s}}{\langle \sigma v \rangle} \right) \left( \frac{\mbox{100 \,\mbox{TeV}}}{m_\sigma}\right)^{3/2},
\ee
where $h$ is the Hubble parameter in units of $100$ km/s/Mpc and we chose some fiducial values with $g_{*} = 10.75$ the number of relativistic degrees of freedom at the time of reheating, $m_\sigma$ is the mass of the decaying scalar, and $m_X$ is the DM mass with interaction rate $\langle \sigma v \rangle$.
We have chosen a fiducial value for the annihilation rate that yields roughly the right amount of DM given a typical set of the parameters (e.g. moduli mass and branching ratio) resulting from the underlying fundamental theory \cite{Acharya:2008bk,Acharya:2009zt}.  And so we see that the cross-section can be as many as three orders of magnitude higher than expected with a thermal history and still give the correct relic density.  We emphasize that the choice of parameters resulting in the correct relic density is not the result of fine-tuning, but instead is robustly motivated by theoretical considerations \cite{Kane:2015jia}. 

{\em Are such scenarios testable?} Indirect detection provides an enormous amount of information by putting upper bounds on the DM annihilation cross section. 
Candidates like non-thermal Winos and Higgsinos with large annihilation cross section are now severely bounded by a combination of continuum and line searches from FERMI data \cite{Fan:2013faa, Cohen:2013ama}. Continuum photons arise from the tree level annihilation process $\chi^0 \chi^0 \rightarrow W^{+} W^{-}$, where $\chi^0$ denotes the WIMP. Dwarf galaxy data rules out Wino DM masses up to around $400$ GeV, while galactic center data rules it out up to around $700$ GeV for either NFW or Einasto profiles \cite{Hryczuk:2014hpa}. These bounds become stronger (weaker) if one considers steeper (core) profiles.  CMB measurements are another source of constraint for non-thermal DM. These constraints arise because 
energy injection from increased DM annihilation can alter the recombination history. We have seen that non-thermal DM can have a larger annihilation rate than thermal WIMPs, and so this would lead to stronger CMB constraints from the additional energy injection. 
The 2015 results from the PLANCK mission \cite{Ade:2015xua} imply a constraint on non-thermal WIMPs with masses 
$\sim 100$ GeV of $f_{\mbox{\tiny eff}} \langle \sigma v \rangle \lesssim 4 \times 10^{-25}$ cm$^3$ s$^{-1}$.  The constant $f_{\mbox{\tiny eff}}$ is a model dependent parameter that depends on the mass of the DM and captures the efficiency of the annihilations -- it typically values of $f_{\mbox{\tiny eff}}$ range from $0.1 -1$.  Although this parameter does introduce 
model dependence and some uncertainty into the constraints, with the 2015 Planck release it is already possible to place reasonable constraints on the annihilation rates of low-mass thermal WIMPs. Before reaching the cosmic variance limit, this type of constraint will be capable of ruling out non-thermal WIMPs completely when they have 
electroweak scale masses and comprise all of the cosmological DM.  Given limited time I have only mentioned a few of the constraints on non-thermal DM and I refer the reader to \cite{Kane:2015jia} for a more comprehensive review including bounds from direct detection and colliders.

\subsection{Matter Power Spectrum and Growth of Structure}
In conventional models for structure formation growth of structure only becomes significant following matter-radiation equality.  This is because DM perturbations $\delta \rho / \rho$ that enter the horizon only grow logarithmically with the scale factor in a radiation dominated universe $\delta \rho / \rho \sim \log a(t) \sim \log t$, whereas they grow linearly in a matter dominated universe $\delta \rho / \rho \sim a(t) \sim t^{2/3}$.  However, in non-thermal cosmologies we have seen that the universe is matter dominated prior to BBN.  This suggests an early period of growth and a new scale at smaller wavelengths in the matter power spectrum.  This possibility was first studied in the absence of annihilations in \cite{Erickcek:2011us} and later more generally in \cite{Fan:2014zua}. It was found that the moduli epoch would introduce a new period of growth for DM perturbations and that the size of the horizon at reheating could provide a new scale for determining the size of the smallest possible DM substructures.  Indeed, for reheat temperatures near the limit set by BBN it was shown that a detectable level of Earth-sized, ultra-compact mini-halos may result.  Of course, whether such structures can survive depends on a number of details, including whether free-streaming and kinetic decoupling effects would destroy the initial formation process. In \cite{Fan:2014zua}, it was found that for typical non-thermal SUSY WIMPs that free-streaming and kinetic decoupling would instead determine the smallest structures.  However, as discussed in \cite{Erickcek:2011us,Fan:2014zua,Erickcek:2015jza} there are exceptions pointing to new directions for model building and a possible signature of the non-thermal period.  One different example appeared in \cite{Erickcek:2015bda}, where it was argued that Bino-like WIMPs could be produced from the radiation present during the non-thermal phase -- so they are thermally produced during the non-thermal period. Thermally produced Binos typically `over-close' the universe because of their feeble interaction rates.  However, when the entropy production occurs near BBN this dilutes the Bino abundance resulting in a viable candidate.  In addition, it was found in \cite{Erickcek:2015bda} that Binos can thermally and kinetically decouple before reheating, and so the enhanced growth of perturbations is preserved introducing a new scale in the matter power spectrum.  If the new structures survive, this leads to an effective `boost factor' (enhanced local densities) increasing the annihilation rate and resulting in $\gamma$-ray fluxes within the reach of FERMI.  That is, the strictly Bino which is difficult to observationally constrain, and typically can't provide the correct relic density of DM, becomes a feasible DM candidate.  Thinking beyond SUSY WIMPs, it would be interesting to explore other DM model building within the non-thermal histories to examine the effect on large scale structure.

\subsection{Testing Inflation with the CMB}
CMB data has reached an impressive level of accuracy and can be used to place stringent constraints on inflation model building and other observables, such as the effective number of relativistic neutrinos $N_{\mbox{\tiny eff}}$ and the acceptable level of isocurvature. In this talk I will discuss briefly the implication of non-thermal histories on testing inflation, but note that there are a number of other important effects which are reviewed in \cite{Kane:2015jia}.

To infer which models of inflation are most likely given the data, we match observations today to the predictions of the inflation model and this matching requires a knowledge of the post-inflationary expansion.  The equation of state (e.g. thermal $w=1/3$ or non-thermal $w=0$) during the post-inflationary epoch determines the expansion rate, and thus the rate at which primordial perturbations (re)enter the Hubble horizon.  For a spectral index $n_s$ that is not strictly scale-invariant ($n_s \neq 1$) this difference in expansion rate will lead to different predictions for the power spectrum \cite{Liddle:2003as}.  
In \cite{Easther:2013nga} the difference between a non-thermal and thermal history was explored and it was shown that the change in equation of state from the thermal case ($w=1/3$) to the moduli phase leads to a shift 
\be \label{deltaNred}
\Delta N=-10.68 + \frac{1}{18}\ln \left[       \left(\frac{g_*(T_r)}{10.75} \right) \left( \frac{T_r}{3 ~\mbox{MeV}} \right)^4 
  \left( \frac{m_p}{\Delta \sigma} \right)^3 \right],
\ee 
where $T_r$ is the reheat temperature, $\Delta \sigma$ is the field displacement, and $g_*$ the number of relativistic degrees of freedom at reheating.
This shift in the number of e-foldings leads to a shift in both the scalar tilt and tensor-to-scalar ratio $n_s$ and $r$, two important parameters for testing models of inflation. 
\bea\label{deltanr}
\nonumber \Delta n_s & = & \left. (n_s-1)\left[-\frac{5}{16}r-\frac{3}{64}\frac{r^2}{n_s-1}\right]\right| \Delta N,\\
\Delta r& = & \left. r \left[(n_s-1)+\frac{r}{8}\right] \right| \Delta N.
\eea 
The resulting fractional  uncertainties  $\Delta r/r, ~\Delta n_s/|n_s-1|$ in these  observables can be substantial \cite{Kinney:2005in,Adshead:2010mc}. For example, the theoretical uncertainty in $n_s$ can be comparable to the precision of observations by Planck \cite{Ade:2015xua}.  

The authors of \cite{Easther:2013nga} argued that accounting for additional consistencies of models, such as requiring a realistic DM candidate, could be used to improve the level of constraint.  As an example, for the non-thermally produced SUSY WIMPs we discussed earlier, constraints from FERMI, PAMELA, and in some cases direct detection (such as Xenon100 and LUX) can place strong constraints on the cross-section.  The authors of \cite{Easther:2013nga} used these constraints to restrict the allowed reheat temperature, which in-turn reduces the uncertainties in $\Delta N$ and so the observables in \eqref{deltanr}.  The bounds on the reheat temperature were later improved in \cite{Cohen:2013ama,Fan:2013faa} with the inclusion of data from HESS and taking into account important effects such as Sommerfeld enhancement. 
Thus, allowing for freedom in the post-inflationary history can lead to less stringent constraints on inflation model building when {\it only} CMB data is taken into consideration, but properly accounting for other components (and the associated data sets) -- in this example we considered the origin and properties of DM -- leads to a new way to begin to probe the post-inflationary epoch.

\section{TOWARD an EFT APPROACH TO REHEATING}
The Effective Field Theory (EFT) of Inflation is based on the idea that there is a physical clock corresponding to the Goldstone boson that non-linearly realizes the spontaneously broken time diffeomorphism (diff)
invariance of the background \cite{Creminelli:2006xe,Cheung:2007st}.  In unitary gauge -- where the clock is homogeneous -- the matter perturbations are encoded within the metric, i.e. the would-be Goldstone bosons are 'eaten' by the metric since  gravity is a gauge theory.   We would like to extend this EFT approach to include inflationary (p)reheating.  

\subsection{Reheating, Broken Symmetries and a Hierarchy of Scales}
Existing investigations of (p)reheating typically rely on choosing a particular potential and then examining the choice of parameters that leads to a successful model.  However, one property that most models have in common is that particle production results during background oscillations of the inflaton field $\phi_(t)$.
A key observation of our approach is that as long as the inflaton dominates the expansion the background will evolve as
\be \label{H_osc}
H(t) = H_{\frw}(t) + H_{\osc}(t) P(\omega t), 
\ee
where the first term dominates at low energies (long wavelengths) and implies an adiabatically evolving, monotonically decreasing contribution to the expansion rate and the second term leads to an oscillatory correction which is sub-leading.
That is, $H_\frw > H_{\osc} P(\omega t)$ with $P(\omega t)$ a quasi-periodic function.
This evolution of the background spontaneously breaks time diffs $t \rightarrow t + \xi^0(t,\vec{x})$ first to a discrete symmetry and then completely.
That is, if we probe the background at energies $E \gg \omega \gg H(t)$, time diffs are realized as a symmetry. 
This remains true until we consider energies comparable to $\omega$. At such energies
$H_{\frw}$ and $H_{\osc}$ will remain nearly constant preserving time diffs,
but the symmetry will be broken by  $P(\omega t)$ to a discrete symmetry $t \rightarrow t + 2\pi \omega^{-1}$. 
At lower energies both 
$H_{\frw}$ and $H_{\osc}$ will also evolve breaking the discrete symmetry.
This symmetry breaking pattern is a natural consequence of the hierarchy of scales that appears in reheating, i.e. high energy (small wavelength) modes probe inflaton oscillations $E / \omega$, whereas low energy (large wavelength) probes capture the expansion of the background $E / H_\frw$ and we have $H_\frw / \omega \ll 1$ during reheating.

%%%%%%%%%%%%%%%%%%%%%%%%%%%%%%%%%%%%%%%%%%%
As an example, consider a simple reheating model where the inflaton oscillates in a potential $V \simeq m^2 \phi^2$.
In this case we can solve for the background evolution 
and one finds \cite{Mukhanov:2005sc}
$H(t)=H_m - {3H_m^2}/({4m}) \sin (2mt+\delta) + \ldots, $
where $H_m = 2/(3t)$ is the Hubble rate in a matter dominated universe and dots represent terms suppressed by higher powers of $H_m/m$.
We see that this is of the form of \eqref{H_osc} corresponding to a matter dominated universe corrected by oscillations suppressed by powers of $H_m/m$.
At energies comparable to the mass of the inflaton we have that the inflaton oscillations break the time diffs, whereas for energies $H \lesssim E \ll m$ the matter dominated expansion 
is primarily responsible for the breaking.  This is another way of stating the familiar fact that on scales comparable to the Hubble radius reheating with a massive inflaton oscillating in a quadratic potential looks like a matter dominated universe, whereas on small scales one can treat the particle production as a local process and in many cases neglect the presence of gravity.

%%%%%%%%%%%%%%%%%%%%%%%%%%%%%%%%%%%%%%%%%%%
\subsection{Inflaton decay, Reheating, and the Role of the Goldstone}
The procedure for constructing the EFT follows analogously to that for inflation \cite{Cheung:2007st,Noumi:2012vr}.
Working in unitary gauge we construct the EFT of fluctuations for reheating in the gravitational and inflationary sectors as

\be
S=\int d^4x \sqrt{-g}\, \left[ \frac{1}{2} m_p^2 R + m_p^2 \dot{H}g^{00} - m_p^2 \left(3H^2+\dot{H} \right) + \frac{M_2^4(t)}{2!} \left( \delta g^{00} \right)^2 +   \frac{M_3^4(t)}{3!} \left( \delta g^{00} \right)^3 + \ldots \right],
\label{theaction}
\ee
where $g^{00}=-1+\delta g^{00}$ and the dots represent terms higher order in fluctuations and derivatives.  
Just as in the inflationary case we now introduce the Goldstone $\pi$ which non-linearly realizes time diffs.    
One of the utilities of the EFT approach is that it is often useful to take a decoupling limit where $\dot{H} \rightarrow 0$ and $m^2_p \rightarrow \infty$ while the combination remains fixed.
For now we focus on operators fixed by tadpole cancelation and take $M_2=M_3=\ldots=0$.  In spatially flat gauge the quadratic action in the decoupling limit is
\be \label{decoupled_action}
S_2=\int d^4x \,a^3 m_p^2 \left[ -\dot{H}\left(\dot{\pi}^2-a^{-2}(\partial_i \pi )^2 \right) -3\dot{H}^2 \pi^2 \right],
\ee
which by canceling the tadpoles has left us with coefficients fixed by the background evolution.
Introducing the canonical field ${\pi}_c =m_p (-2 \dot{H})^{1/2} {\pi}$ one can show that the term responsible for breaking the shift symmetry is due to the oscillatory behavior of the time-dependent potential of the inflaton (corresponding to an operator $\hat{{\cal O}}_\pi \sim V^{\prime \prime} \pi_c^2$) and does {\it not} come from mixing with gravity.
As in the EFT of Inflation the leading mixing with gravity scales as $E_{\mbox{\tiny mix}}=\epsilon^{1/2} H=\dot{H}^{1/2}$, although a difference for us is given $V\sim \phi_0^n$ then $\epsilon=3n/(n+2)$ is typically an order one number.
The decoupling limit will be useful for probing scales with $E \gg E_{\mbox{\tiny mix}}$, but other times it is appropriate to include corrections coming from the mixing with gravity.
One useful aspect of \eqref{decoupled_action} is to study the stability of sub-horizon perturbations against collapse. Just as in studies of ghost condensation \cite{ArkaniHamed:2003uy}, including higher corrections to the EFT (e.g. $M_2 \neq 0$) could lead to new and consistent models for (p)reheating.
Part of the utility of our approach is that inflaton self-interactions will also be fixed by the symmetries. For (p)reheating this implies that if one is interested in interactions, which determine rescattering and backreaction effects, the coefficients for these terms that appear in the action will also be fixed by the same symmetries. We postpone further discussion of this for the future and instead turn to 
the issue of particle production of the reheat field.

\subsection{Reheating Fields and Coupling to the Inflaton}
Given the EFT description we now couple the inflationary sector to an additional reheat field $\sigma(t,\vec{x})$.
We will be interested in the production of $\sigma(t,\vec{x})$ particles resulting from the oscillations of the background inflaton field $\phi_0(t)$.
In unitary gauge, the production of particles by the background will result from operators $f(t) \hat{{\cal O}}_n(\sigma)$.
At the quadratic level this gives
\be
S^{(2)}_{\sigma} =\int d^4x\sqrt{-g}\left[-\frac{\alpha_1(t)}{2}g^{\mu\nu}\partial_\mu\sigma\partial_\nu\sigma 
+\frac{\alpha_2(t)}{2}(\partial^o\sigma)^2-\frac{\alpha_3(t)}{2}\sigma^2+\alpha_4(t)\sigma\partial^o\sigma\right], 
\label{sigma_action}
\ee
where we see that the broken time diffs allow for a non-trivial sound speed $c_\sigma^2=\alpha_1/(\alpha_1+\alpha_2)$.
The action \eqref{sigma_action} already accounts for many existing models in the literature.  For example, preheating with $V \sim g^2 \phi_0^2 \sigma^2$ corresponds to 
$\alpha_1=1,\alpha_2=\alpha_4=0$ and $\alpha_3=g^2 \phi_0(t)^2$.  Whereas, if we require the inflaton to remain shift symmetric throughout reheating, 
as one might anticipate in models of Natural Inflation, then we consider interactions of the form $(\partial_\mu \phi_0)^2 \sigma^2/\Lambda^2$,
where $\Lambda$ is the cutoff for the {\it background}.
Our approach captures this model by now choosing $\alpha_3=2\dot{\phi}_0(t)^2/\Lambda^2$. 
It has been found that preheating in models that preserve an inflaton shift symmetry is not efficient \cite{ArmendarizPicon:2007iv}.
One reason for this is that naively we assume that the energy of the fields can not exceed the cutoff $\Lambda$.
However, an advantage of our EFT approach is that the parameters, such as
$\alpha_3$, can be completely non-linear and their origin is irrelevant since the background itself is not physically observable.
This is analogous to the EFT of Inflation, where noting that the background is not an observable the authors assume, a priori, a quasi-de Sitter background and then study the EFT
of fluctuations about that background.

Although the parameters are not directly observable, we can constrain the EFT parameters in several ways.  
Just as in the case of the inflaton above, avoiding instabilities will place constraints.
Moreover, we must require that the coefficients violate adiabaticity so that particles are produced --
this implies that $\dot{\alpha}_3 / \alpha^2_3 \gg1$. 
When adiabaticity fails $\sigma$-quanta will be produced.  A calculation of the precise number of particles that results from the violation 
would require a detailed investigation and would seem to still be model dependent.
However, an advantage of our approach is that these same coefficients will enter into interaction terms with their form dictated by non-linearly realizing time diffs.
Interactions between inflaton particles ($\delta \phi \sim \pi$) and the reheat particles ($\sigma$) will have two origins in the EFT.
Firstly, upon reintroducing the Goldstone field via $t \rightarrow t+\pi$ the coefficients in \eqref{sigma_action} will generate interactions.  Secondly,
there will be mixing terms of the form $\sim (\delta g^{00})^m \hat{{\cal O}}_n(\sigma)$. As an example of the first type there is an operator $\sim \dot{\alpha}_3 \pi \sigma^2$ 
and so we see when adiabaticity is violated $\dot{\alpha}_3 \gg  \alpha_3^2$ this interaction would become important.
Details will appear in a future publication, here we just want to emphasize that the symmetries will enforce connections like this and so 
our approach connects successful particle production (violations of adiabaticity) with the interactions that are important for understanding
particle back-reaction, rescattering, and the details of thermalization.

\subsection{Concluding thoughts}
Here I have outlined a program to establish a model independent approach to (p)reheating.
We have seen that requiring the Goldstone to non-linearly realize the broken time diffeomorphisms implies important connections
between the background dependent parameters of the EFT. In particular, we have seen that the amount of particle production and the sound speed
is related to the strength of interactions which are important for understanding the duration of (p)reheating and the thermalization process.  

There are some challenges for our approach.  We have assumed initially that the reheat field does not substantially contribute to the energy density.
This assumption is certainly justified during the first phases of preheating \cite{Kofman:1997yn}, however as particles continue to be produced they may influence the duration of reheating.
In the case of the EFT of inflation this can cause an ambiguity in the (clock) Goldstone description \cite{LopezNacir:2011kk}.  Here we have argued that by tracking the adiabatic mode and neglecting 
any isocurvature we can avoid this problem, but this issue requires a more careful investigation.  Another issue is that unlike the EFT of inflation where primordial non-gaussianity (or lack of it) was an important observation for restricting the parameters of the EFT, such observations seem rather irrelevant for (p)reheating.  Instead, aside from the possibility of gravity wave signatures, our main application is developing a framework to analytically understand the thermalization process. As we have seen above, the form and strength of these interactions are restricted by the symmetries. This, along with the model independence that results from our treatment of the background suggests this is a promising approach to further develop.

% Acknowledgement
\section{ACKNOWLEDGMENTS}
I would like to thank Adrienne Erickcek, JiJi Fan, Gordy Kane, Ogan Ozsoy, Marco Peloso, Gizem Sengor, and Masahide Yamaguchi for collaboration and/or discussions. I am especially grateful to Kuver Sinha for many significant contributions and discussions that led to much of the research I am reporting here. This research was funded in part by NASA Astrophysics Theory Grant NNH12ZDA001N, DOE grant DE-FG02-85ER40237, and National Science Foundation Grant No. PHYS-1066293 and the hospitality of the Aspen Center for Physics. I would also like to thank the organizers of PPC 2015 and especially Barbara Szczerbinska for a wonderful venue and a great conference.

%\nocite{*}
\bibliographystyle{aipnum-cp}%

\end{document}